# Tri-level Scheduling Model Considering Residential Demand Flexibility of Aggregated HVACs and EVs under Distribution LMP

Xiaofei Wang, *Student Member, IEEE,* Fangxing Li, *Fellow, IEEE,* Jin Dong, *Member, IEEE,*
Mohammed. Olama, *Senior Member, IEEE,* Qiwei Zhang, *Student Member, IEEE,*
Qingxin Shi, *Member, IEEE,* Byungkwon Park, *Member, IEEE,* Teja Kuruganti, *Senior Member, IEEE*

*Abstract*— Residential loads, especially heating, ventilation and air conditioners (HVACs) and electric vehicles (EVs), have great potentials to provide demand flexibility which is an attribute of Grid-interactive Efficient Buildings (GEB). Under this new paradigm, first, EV and HVAC aggregator models are developed in this paper to represent the fleet of GEBs, in which the aggregated parameters are obtained based on a new approach of data generation and least squares parameter estimation (DG-LSPE), which can deal with heterogenous HVACs. Then, a tri-level bidding and dispatching framework is established based on competitive distribution operation with distribution locational marginal price (DLMP). The first two levels form a bilevel model to optimize the aggregators' payment and to represent the interdependency between load aggregators and the distribution system operator (DSO) using DLMP, while the third level is to dispatch the optimal load aggregation to all residents by the proposed priority list-based demand dispatching algorithm. Finally, case studies on a modified IEEE 33-Bus system illustrates three main technical reasons of payment reduction due to demand flexibility: load shift, DLMP step changes, and power losses. They can be used as general guidelines for better decision making for future planning and operation of demand response programs.

*Index Terms*— EV aggregator, HVAC aggregator, distribution locational marginal price (DLMP), residential demand flexibility, tri-level scheduling model, load shift, DLMP step change.

## NOMENCLATURE

*Sets*
| | |
|---|---|
| $T$ | Set of time slots |
| $H$ | Set of HVAC aggregators |
| $E$ | Set of EV aggregators |
| $G$ | Set of all generators |
| $B$ | Set of all nodes |
| $Sub$ | Substation |

*Constants*
| | |
|---|---|
| $P_{i,t}^{MT,min}$ | Minimum active power of MT $i$ at time $t$ |
| $P_{i,t}^{MT,max}$ | Maximum active power of MT $i$ at time $t$ |
| $\alpha_i$ | Power factor of DG $i$ |
| $P_{i,t}^{PV,min}$ | Minimum active power of PV $i$ at time $t$ |
| $P_{i,t}^{PV,max}$ | Maximum active power of PV $i$ at time $t$ |
| $Q_{i,t}^{SVC,min}$ | Minimum reactive power of SVC $i$ at time $t$ |
| $Q_{i,t}^{SVC,max}$ | Maximum reactive power of SVC $i$ at time $t$ |
| $\eta^C, \eta^D$ | Charge/discharge efficiency of EV |
| $P_{i,t}^{C,min}$ | Minimum charge power of EV aggregator $i$ |
| $P_{i,t}^{Dis,max}$ | Maximum discharge power of EV aggregator $i$ |
| $SOC_i^{E,min}$ | Minimum state of charge of EV aggregator $i$ |
| $SOC_i^{E,max}$ | Maximum state of charge of EV aggregator $i$ |
| $E_i^R$ | Rated energy capacity of EV aggregator $i$ |
| $N_i^e$ | Number of EVs participating in aggregator $i$ |
| $N_i^{piles}$ | Number of charging piles in charging station $i$ |
| $p_j^c$ | Rated charging power of EV $j$ |
| $E_j^r$ | Rated energy of EV $j$ |
| $e_j^d$ | Driving consumption of EV $j$ per mile |
| $d_j$ | Driving distance over a day |
| $a_j, b_j, g_j$ | Coefficients of the thermal function of HVAC $j$ |
| $\theta_{out}$ | Day-ahead forecasted outdoor temperature |
| $\theta^{min}, \theta^{max}$ | Comfortable temperature boundary |
| $R_j, C_j$ | Thermal resistance and capacitance of HVAC $j$ |
| $\eta$ | Cooling efficiency of HVAC |
| $P^{rated}$ | Rated power of HVAC |
| $\tilde{a}_i, \tilde{b}_i, \tilde{g}_i$ | Coefficients of the thermal transfer function of HVAC aggregator $i$ |
| $N_i^h$ | Number of HVACs participating in aggregator $i$ |
| $syn_i^{min}, syn_i^{max}$ | Minimum/maximum synchronicity rate of HVAC aggregator $i$ |
| $\Delta u_i^{dr}, \Delta u_i^{ur}$ | Ramp up/down rate of HVAC aggregator $i$ |
| $SOC_i^{H,min}$ | Minimum state of charge of HVAC aggregator $i$ |
| $SOC_i^{H,max}$ | Maximum state of charge of HVAC aggregator $i$ |
| $c_{i,t}, d_{i,t}$ | Active/reactive generation cost of generator $i$ |
| $P_{i,t}^D, Q_{i,t}^D$ | Fixed active and reactive load demand of node $i$ at time $t$ |

*Variables*
| | |
|---|---|
| $P_{i,t}^{MT}, P_{i,t}^{MT}$ | Active/reactive power of MT $i$ at time $t$ |
| $P_{i,t}^{PV}, P_{i,t}^{PV}$ | Active/reactive power of PV $i$ at time $t$ |
| $Q_{i,t}^{SVC}$ | Reactive power of SVC $i$ at time $t$ |
| $E_{i,t}$ | Energy of EV aggregator $i$ at time $t$ |
| $P_{i,t}^C, P_{i,t}^{Dis}$ | Charge/discharge power of EV aggregator $i$ |
| $\theta_{j,t}$ | Indoor temperature of building $j$ at time $t$ |
| $u_{j,t}$ | Binary variable stating the ON/OFF of HVAC $j$ |
| $\tilde{\theta}_{i,t}$ | Equivalent indoor temperature of HVAC aggregator $i$ |
| $\tilde{u}_{i,t}$ | ON-state ratio of HVACs in aggregator $i$ at time $t$ |
| $P_{i,t}^H$ | Active power of HVAC aggregator $i$ at time $t$ |
| $SOC_{i,t}$ | State of charge of HVAC aggregator $i$ at time $t$ |
| $\pi_{i,t}^p$ | Active DLMP of node $i$ at time $t$ |
| $P_t^{loss}, Q_t^{loss}$ | Active/reactive power loss at time $t$ |
| $V_{j,t}$ | voltage of node $j$ at time $t$ |

## I. INTRODUCTION

INSPIRED by the smart grid concept, at the generation side, the deployment of distributed generators (DGs), such as photovoltaic (PV), microturbine (MT) and wind turbine (WT), has been increasing in the past decades in distribution systems [1]. Meanwhile, at the load side, industrial and commercial customers are encouraged to participate in demand response (DR) programs. The proliferation of all these types of distributed energy resources (DERs) makes the distribution system more flexible and active [2]. Also, the advanced metering infrastructure like bilateral smart meter facilitates the information exchange between DERs and the distribution system operator (DSO) [3]. With this background, it is believed that DERs are driving the transition from a passive distribution system to a market-based one that aims to achieve the optimal allocation of all DERs [4].

In the transmission level market, locational marginal price (LMP) has been widely implemented by ISOs, such as PJM, New York ISO, ISO-New England, etc. [5]. In the research community, there are some previous works extending LMP to distribution locational marginal price (DLMP). In [6], DLMP is developed in a distribution system. In [7] [8], DSO determines DLMP based on generation offers and load bids by clearing the market. Refs. [2] and [9] integrate the voltage component for DLMP and LMP, respectively. Ref. [10] provides an interval prediction for the DLMP considering the uncertainty of renewables. All these works show that DLMP can reflect real electricity price information in distribution. Thus, DLMP can be naturally considered as a price signal to incentivize consumers to adjust their loads to save electricity bills.

According to [11], heating, ventilation and air conditioners (HVACs) account for 45% of average summer peak-day loads. Also, building's characteristic of thermal storage provides great demand flexibility by shedding and shifting HVAC load because indoor temperature does not change fast due to thermal inertia [12][13]. Together with electric vehicles (EVs) that have electricity storages, they are the ideal residential DR candidates to provide demand flexibility which is an attribute of Grid-interactive Efficient Buildings (GEB).

Appropriate price signals can efficiently guide consumers to change their consumption patterns. This is both beneficial to the consumers and the distribution system. Previous studies of the DR with HVACs and EVs in response to dynamic prices can be categorized into two groups based on with or without market. 1) Without market environment: [14] studies the optimal precooling of HVACs under time varying electricity prices. [15] proposes a dynamic DR control strategy to adjust the set-point temperature of HVACs according to real time prices. [16] presents an alternating direction method of multipliers (ADMM)-based residential DR management strategy. 2) With market environment: in [17][18], DLMP is utilized to optimize the EV charging schedule to alleviate the congestion issue. In [19][20], DLMP considering distribution congestion price is proposed to guide the DR to prevent congestion. These studies show the effectiveness of DLMP in congestion management. However, the power losses are usually too high to be neglected due to the high R/X ratio, and the voltage is critical for the reliable operation of a distribution system, DLMP should be able to reflect these costs. In addition, compared with the broad DR concept, proper integration with the actual HVAC/EV models is highly necessary for industrial deployment. Also, as a participant in a competitive market, residential loads integrated at a large scale can affect DLMP. Thus, residents have the motivation to consume electricity strategically to minimize their electricity bills, possible via an aggregator.

Based on these considerations, we focus on reducing residents' electricity bills by proposing a novel tri-level model based on DLMP. The first level is to minimize residents' electricity payment by optimizing aggregated residential DR schedules according to DLMP in the second level. The second level is to clear the day-ahead distribution market and integrate power losses and voltage constraints in DLMP. Since the first two levels are coupled, the bilevel model is solved by reformulating it as a single level mathematical programming with equilibrium constraints (MPEC) by Karush-Kuhn-Tucker (KKT) optimality conditions and then a mixed-integer linear programming (MILP) by the big-M method. Once the optimal aggregated HVAC/EV schedule is obtained, the 3rd level is solved to dispatch all DR-participating residents based on the aggregated schedule. The main contributions are as follows:

• The HVAC aggregator and EV aggregator models are developed on behalf of end users. In particular, a data generation and least-square parameter estimation (DG-LSPE) algorithm is applied to aggregate heterogenous HVACs, which differs from previous work assuming HVACs are homogenous.

• In a competitive distribution market with DLMP, a tri-level model is proposed for aggregators to minimize their power purchasing payment and maintain the secure and economic operation of the distribution system.

• As illustrated in the case studies, the economic benefits of residential demand flexibility can be attributed to three reasons: load shifts from high-cost hours to low-cost hours, DLMP step changes, and power loss reductions.

The rest of this paper is organized as follows. Section II describes the three-layer day-ahead distribution market structure and market participants including EV aggregator, HVAC aggregator and DGs. Section III proposes the tri-level optimization model and presents the solution method. Case studies in Section IV demonstrate the effectiveness of the proposed model and analyze both economic and operational benefits of flexible demand. Section V concludes the paper.

## II. STRUCTURE AND PARTICIPANTS OF DISTRIBUTION-LEVEL ELECTRICITY MARKET

This section describes the structure of distribution-level electricity market and the models of different participants.

### A. Market Structure

In general, an individual resident is not eligible to participate in the market directly due to the complex market rules, strict participation requirements [21], and heavy calculation burden [16]. To address these challenges, aggregators have emerged to serve as intermediaries between these residents and DSO to provide demand flexibility. After collecting comfortable temperature zones and charging requirements, aggregators help end residents respond to DLMP by bidding in the distribution market and dispatching the optimal power. As such, the structure of the three-layer distribution-level electricity market is illustrated in Fig. 1.

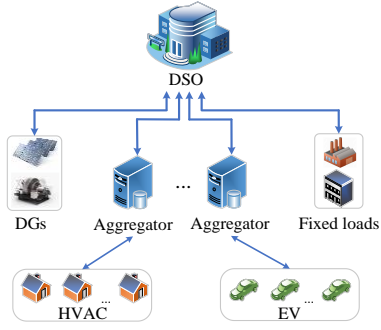

Fig. 1. Three-layer day-ahead distribution-level electricity market

From Fig. 1, DGs like PV and MT participate in the market by submitting their offers to the DSO. The loads are divided into fixed load without price elasticity and flexible load that responses to DLMP, which refers to HVAC and EV load in this paper. Fixed loads are only price takers and they are forecasted according to historical data. HVAC and EV aggregators submit their load quantities and let other participants decide the electricity price. The DSO clears the day-ahead market and broadcast DLMP to the whole distribution system.

*B. EV Aggregator*

For the sake of simplicity, we assume EV aggregators are located at the same nodes with charging stations, and there are limited charging piles in each charging station. Thus, aggregated EV charging and discharging power are limited by this constraint. Here, EV aggregator is modeled as the summation of all EVs.

$$E_{i,t+\Delta t} = E_{i,t} + \Delta t \left( \eta^C P_{i,t}^C - 1/\eta^D P_{i,t}^{Dis} \right) \quad (1)$$

$$SOC_i^{E,\min} \cdot E_i^R \leq E_{i,t+\Delta t} \leq SOC_i^{E,\max} \cdot E_i^R \quad (2)$$

$$0 \leq P_{i,t}^C \leq P_i^{C,\max} \quad (3)$$

$$E_{i,t=0} \leq E_{i,t=T} \quad (4)$$

$$\sum_{t=1}^{T} \Delta t \cdot P_{i,t}^{Dis} = \sum_{j=1}^{N_i^e} e_j^d \cdot d_j \quad (5)$$

where Eq. (1) represents the energy of EV aggregator $i$, assuming all EVs have the same charging and discharging efficiency, (2) represents the state of charge (SOC) limits, (3) is the charging power limits, (4) ensures the total charging energy is no less than the total discharging energy over one day, (5) describes the driving consumption constraints. $P_i^{C,\max}$, $E_i^R$ and $E_{i,t=0}$ are obtained as follows.

$$P_i^{C,\max} = \sum_{j=1}^{N_i^{piles}} p_j^C \quad (6)$$

$$E_i^R = \sum_j^{N_i^e} E_j^r \quad (7)$$

$$E_{i,t=0} = \sum_j^{N_i^e} E_{j,t=0} \quad (8)$$

*C. HVAC Aggregator*

*1) Single HVAC:* For simplicity, the first order thermal transfer function is utilized to model a building's dynamic indoor temperature [12]. Thus, each HVAC in cooling mode can be modeled with the following equations:

$$\theta_{j,t+\Delta t} = a_j \theta_{j,t} + b_j \theta_{out,t} + g_j u_{j,t} \quad (9)$$

$$\theta^{\min} \leq \theta_{j,t+\Delta t} \leq \theta^{\max} \quad (10)$$

where $a_j = 1 - \Delta t/C_j R_j$, $b_j = \Delta t/C_j R_j$, $g_j = -\eta P^{rated} \Delta t/C_j$.

The typical parameters of a single EV and HVAC are shown in TABLE I.

TABLE I. PARAMETERS OF SINGLE EV AND HVAC

| EV | Parameter | HVAC | Parameter |
|---|---|---|---|
| $E_j^r$ | $N(50, 1^2)$ (kWh) | $P^{rated}$ | 5 (kW) |
| $p_j^C$ | $N(7.2, 0.2^2)$ (kW) | $R_j$ | $N(R_A, 0.2^2)$ (°C/kW) |
| $e_j^d$ | $N(0.23, 0.01^2)$ (kWh/mile) | $C_j$ | $N(C_A, 0.2^2)$ (kWh/°C) |
| $\eta^C/\eta^D$ | 0.98 | $\eta$ | 3 |
| $E_{j,t=0}$ | $E_j^r \cdot U[0.2, 0.3]$ (kWh) | $\theta^{\max}$ | 21 (°C) |
| $d_j$ | $U[30, 50]$ (miles) | $\theta^{\min}$ | 19 (°C) |
| $\Delta t$ | 1 (hour) | $\Delta t$ | 1 (hour) |

*2) HVAC aggregator:* Different from the previous analytical HVAC aggregation which assumes homogenous collection of HVACs [22], this paper supposes that HVACs are heterogeneous to ensure the diversity of individual buildings where $R_j$ and $C_j$ of all buildings follow a given probabilistic distribution. Note, a normal distribution is applied here although any distribution will work, because we simply need to generate data for aggregation. Then, a data-driven approach is proposed to estimate the aggregated parameters. After that, the HVAC aggregator's thermal equation is modeled and the corresponding operation constraints are presented below.

$$\tilde{\theta}_{i,t+\Delta t} = \tilde{a}_i \tilde{\theta}_{i,t} + \tilde{b}_i \theta_{out,t} + \tilde{g}_i \tilde{u}_{i,t} \quad (11)$$

$$P_{i,t}^H = \tilde{u}_{i,t} \cdot N_i^h \cdot P^{rated} \quad (12)$$

$$\theta^{\min} \leq \tilde{\theta}_{i,t+\Delta t} \leq \theta^{\max} \quad (13)$$

$$syn_i^{\min} \leq \tilde{u}_{i,t} \leq syn_i^{\max} \quad (14)$$

$$-\Delta u_i^{dr} \leq \tilde{u}_{i,t+\Delta t} - \tilde{u}_{i,t} \leq \Delta u_i^{ur} \quad (15)$$

$$SOC^{H,\min} \leq SOC_{i,t} \leq SOC^{H,\max} \quad (16)$$

where Eq. (11) represents the dynamic temperature of the HVAC aggregator, $\tilde{\theta}_{i,t} = \sum_{j=1}^{N_i^h} \theta_{j,t} / N_i^h$, $\tilde{u}_{i,t} = \sum_{j=1}^{N_i^h} u_{j,t} / N_i^h$, $\tilde{u}_{i,t}$ is a continuous variable in [0, 1], (12) is to obtain the active power of HVAC aggregator, (14) is synchronicity constraint which limits the number of HVACs to be turned ON at the same period, (15) is ramp up/down constraint to limit the state transfer speed of HVACs, (16) is energy constraint to reduce the probability of all HVACs centering at the temperature boundaries [23] in order to improve the load dispatching performance. The $SOC_{i,t}$ is formulated as

$$SOC_{i,t} = \frac{\sum_{j=1}^{N_i^h} (\theta_{\max} - \theta_{j,t})}{N_i^h (\theta_{\max} - \theta_{\min})} = \frac{\theta_{\max} - \tilde{\theta}_{i,t}}{\theta_{\max} - \theta_{\min}} \quad (17)$$

*3) DG-LSPE:* HVAC control strategy plays an important role in the simulation of HVACs. Priority list control [24], a direct and effective control strategy, is utilized to control the state transformation of HVACs in this paper. Based on this, a DG-LSPE algorithm is designed to determine $\tilde{a}_i$, $\tilde{b}_i$ and $\tilde{g}_i$. Because every HVAC aggregator uses the same estimation algorithm, subscript $i$ is neglected and the algorithm is described below.

**Algorithm 1. DG-LSPE**

**Step 1. Generate a data set:** Simulate the indoor temperatures of $N_h$ buildings in a whole day at different $\tilde{u}$.

1. Generate $R_j$ and $C_j$ of $N_h$ buildings;
2. For $\tilde{u} = 0.1 : \Delta u : 1$;

3. For $t = 1:\Delta t:T$;
4. Obtain the number of ON state devices: $N_{on} = N_h \cdot \tilde{u}$;
5. Choose the top $N_{on}$ devices to form the turned-ON priority list according to the descending order of indoor temperature, and the rest devices form the turned-OFF list;
6. Calculate the indoor temperature for every building according to Eq. (9) by carrying out the turned-ON and turned-OFF list;
7. Calculate the average indoor temperature, and record input data $[\tilde{\theta}_t, \theta_{out,t}, \tilde{u}]$ and output data $\tilde{\theta}_{t+\Delta t}$;
8. End $t$;
9. End $\tilde{u}$.

**Step 2. Least squares parameter estimation:**
Obtain the aggregated parameters by the least squares estimation.

$$x = (A^T A)^{-1} A^T c$$

where $A$ and $c$ both have $(T/\Delta t\text{ -}1)\cdot(1/\Delta u)$ rows and

$$Ax = c \text{ where } A_i = \begin{bmatrix} \tilde{\theta}_1 & \theta_{out,t} & \tilde{u}_i \\ \vdots & \vdots & \vdots \\ \tilde{\theta}_{T/\Delta t-1} & \theta_{out,T/\Delta t-1} & \tilde{u}_i \end{bmatrix}, c_i = \begin{bmatrix} \tilde{\theta}_2 \\ \vdots \\ \tilde{\theta}_{T/\Delta t} \end{bmatrix}$$

$$A = \begin{bmatrix} A_1 \\ \vdots \\ A_{(1/\Delta u)} \end{bmatrix}, x = \begin{bmatrix} \tilde{a} \\ \tilde{b} \\ \tilde{g} \end{bmatrix}, c = \begin{bmatrix} c_1 \\ \vdots \\ c_{(1/\Delta u)} \end{bmatrix}$$

### D. DG Models

According to the active or reactive power output and control strategy, most DGs fall into synchronous machines, inverter-based machines, and var compensation devices. For simplicity, three typical representatives are included in this paper.

MT is a synchronous machine-based DG. Its output power should satisfy the physical constraints and the power factor requirement [2]. In this paper, we set $\alpha_i$ to 0.95.

$$P_{i,t}^{MT,\min} \leq P_{i,t}^{MT} \leq P_{i,t}^{MT,\max} \quad (18)$$

$$0 \leq Q_{i,t}^{MT} \leq P_{i,t}^{MT} \tan(\arccos \alpha_i) \quad (19)$$

PV is an inverter-based generator that can both absorb or generate reactive power. The power factor range is $[\alpha_i$ lagging, $\alpha_i$ leading].

$$P_{i,t}^{PV,\min} \leq P_{i,t}^{PV} \leq P_{i,t}^{PV,\max} \quad (20)$$

$$-P_{i,t}^{PV} \tan(\arccos \alpha_i) \leq Q_{i,t}^{PV} \leq P_{i,t}^{PV} \tan(\arccos \alpha_i) \quad (21)$$

Static Var compensator (SVC) is induced to help maintain the voltage due to three attributes: 1) can generate or absorb reactive power, 2) output power can be adjusted continuously, 3) a good tradeoff between cost and performance [2].

$$Q_{i,t}^{SVC,\min} \leq Q_{i,t}^{SVC} \leq Q_{i,t}^{SVC,\max} \quad (22)$$

## III. TRI-LEVEL MODEL

As the representative of residential users, an aggregator's motivation is to optimize its financial benefit. The aggregator has two tasks: 1) bid strategically in the distribution market to obtain the optimal schedule, and 2) identify the optimal allocation of this schedule to each residential user who may have HVAC and EV loads. For 1), the aggregators and the DSO are different entities with different interests, meanwhile, an aggregator's demand and the system DLMP are coupled variables, which means any change in one variable will cause changes in the other one and vice versa. Therefore, a bilevel model is formulated to represent this relationship. For 2), for simplicity, we set the aggregator and residents follow a vertical dispatcher-receiver structure. That is, after the aggregator's total load demand is determined, the third level model will allocate the total demand to all participating residents.

### A. First Level

The first level is to minimize aggregator's power purchasing cost while satisfying comfort and charging requirements.

$$\min \sum_{t \in T} \left( \sum_{i \in H} \pi_{i,t}^p \cdot P_{i,t}^H + \sum_{i \in E} \pi_{i,t}^p \cdot P_{i,t}^C \right) \quad (23a)$$

$$s.t. \text{ constraints } (1)\text{-}(17) \quad (23b)$$

### B. Second Level

The second level is a market clearing model which has the objective to minimize system generation cost while maintaining all operational constraints. Note that the power congestion constraint is neglected in this study because congestion rarely happens in real distribution systems with radial topologies.

$$\min \sum_{t \in T} \sum_{i \in G} c_{i,t} \cdot P_{i,t}^G + d_{i,t} \cdot \hat{Q}_{i,t}^G \quad (24a)$$

$$\text{where } G = \{Sub, PV, MT, SVC\}$$

s.t.

$$\sum_{i \in G} P_{i,t}^G - \sum_{i \in B} P_{i,t}^D - \sum_{i \in H} P_{i,t}^H - \sum_{i \in E} P_{i,t}^C - P_t^{loss} = 0 : \lambda_t^p, \forall t \in T \quad (24b)$$

$$\sum_{i \in G} Q_{i,t}^G - \sum_{i \in B} Q_{i,t}^D - Q_t^{loss} = 0 : \lambda_t^q, \forall t \in T \quad (24c)$$

$$V_{j,t} = V_{1,t} + \sum_{i \in B} Z_{j,i}^p \left( P_{i,t}^G - P_{i,t}^D - P_{i,t}^H - P_{i,t}^C \right) + \sum_{i \in B} Z_{j,i}^q \left( Q_{i,t}^G - Q_{i,t}^D \right) \quad (24d)$$

$$V^{\min} \leq V_{j,t} \leq V^{\max} : \omega_{j,t}^{v,\min}, \omega_{j,t}^{v,\max}, \forall j \in B, \forall t \in T \quad (24e)$$

$$P_{i,t}^{G,\min} \leq P_{i,t}^G \leq P_{i,t}^{G,\max} : \omega_{i,t}^{p,\min}, \omega_{i,t}^{p,\max}, \forall i \in Sub, \forall t \in T \quad (24f)$$

$$(18)(20): \omega_{i,t}^{p,\min}, \omega_{i,t}^{p,\max}, \forall i \in \{MT, PV\}, \forall t \in T \quad (24g)$$

$$Q_{i,t}^{G,\min} \leq Q_{i,t}^G \leq Q_{i,t}^{G,\max} : \omega_{i,t}^{q,\min}, \omega_{i,t}^{q,\max}, \forall i \in Sub, \forall t \in T \quad (24h)$$

$$(19)(21)(22): \omega_{i,t}^{q,\min}, \omega_{i,t}^{q,\max}, \forall i \in \{MT, PV, SVC\}, \forall t \in T \quad (24i)$$

$$-Q_{i,t}^G \leq \hat{Q}_{i,t}^G, Q_{i,t}^G \leq \hat{Q}_{i,t}^G : \kappa_{i,t}^-, \kappa_{i,t}^+, \forall i \in G, \forall t \in T \quad (24j)$$

where (24b) and (24c) represent the active and reactive power balance constraints, the substation is regarded as a generator with a large capacity; (24d) is the voltage expression derived from linearized power flow for distribution (LPF-D) [7], in which $Z^p$ and $Z^q$ are matrices of nodal voltage change with respect to net power injections that can be derived from [7]; (24e) is the voltage limits; Node 1 is the reference bus that connects to the substation; (24f) - (24i) are generators' active and reactive power output limits; in (24j), $\hat{Q}_{i,t}^G = |Q_{i,t}^G|$ since both absorbing and generating reactive power will induce cost [2]. The power loss and power loss factor are obtained by the loss factors for distribution (LF-D) in [7], then power loss is linearized by Taylor's series:

$$P_t^{loss} \approx P_t^{loss*} + \sum_{i \in B} \frac{\partial P_t^{loss}}{\partial P_{i,t}^G} \left( \Delta P_{i,t}^G - \Delta P_{i,t}^D \right) + \sum_{i \in B} \frac{\partial P_t^{loss}}{\partial Q_{i,t}^G} \left( \Delta Q_{i,t}^G - \Delta Q_{i,t}^D \right) \quad (25)$$

$$Q_t^{loss} \approx Q_t^{loss*} + \sum_{i \in B} \frac{\partial Q_t^{loss}}{\partial P_{i,t}^G} \left( \Delta P_{i,t}^G - \Delta P_{i,t}^D \right) + \sum_{i \in B} \frac{\partial Q_t^{loss}}{\partial Q_{i,t}^G} \left( \Delta Q_{i,t}^G - \Delta Q_{i,t}^D \right) \quad (26)$$

where $\Delta P_{i,t}^G = P_{i,t}^G - P_{i,t}^{G*}$ represents the power difference between two close operating points, and $\Delta Q_{i,t}^G$, $\Delta P_{i,t}^D$ and $\Delta Q_{i,t}^D$ have similar expressions.

The Lagrange function can be written as follows.

$$L = \sum_{t \in T} \sum_{i \in G} c_{i,t} \cdot P_{i,t}^G + d_{i,t} \cdot \hat{Q}_{i,t}^G$$
$$- \sum_{t \in T} \lambda_t^p \left( \sum_{i \in G} P_{i,t}^G - \sum_{i \in B} P_{i,t}^D - \sum_{i \in H} P_{i,t}^H - \sum_{i \in E} P_{i,t}^C - P_t^{loss} \right)$$
$$- \sum_{t \in T} \lambda_t^q \left( \sum_{i \in G} Q_{i,t}^G - \sum_{i \in B} Q_{i,t}^D - Q_t^{loss} \right)$$
$$- \sum_{t \in T} \sum_{j \in B} \omega_{j,t}^{v,\min} \left( V_{j,t} - V^{\min} \right) - \sum_{t \in T} \sum_{j \in B} \omega_{j,t}^{v,\max} \left( V^{\max} - V_{j,t} \right)$$
$$- \sum_{t \in T} \sum_{i \in G} \omega_{i,t} \cdot g_{i,t}(x) \quad (27)$$

where $g_{i,t}(x)$ represents the power output limits in (24f) - (24j).

The active DLMP is the first order partial derivative of the Lagrange function with respect to the active power. According to Eq. (28), the active DLMP can be expressed as:

$$\pi_{i,t}^p = \frac{\partial L}{\partial P_{i,t}^D} = \lambda_t^p + \lambda_t^p \cdot \frac{\partial P_t^{loss}}{\partial P_{i,t}^D} + \lambda_t^q \cdot \frac{\partial Q_t^{loss}}{\partial P_{i,t}^D}$$
$$+ \sum_{j \in B} \left( \omega_{j,t}^{v,\min} - \omega_{j,t}^{v,\max} \right) Z_{j,i}^p \quad (28)$$

### C. Third Level

The third level is to dispatch the optimal load demand for all individual end residents. Based on priority list control, the load dispatching algorithm for HVACs, with a 10-min time step (i.e., 6 time slots in one hour) to achieve more accurate control, is presented in **Algorithm 2**.

**Algorithm 2**. HVAC load dispatching
1. For $t = 1:1: T$;
2. For $k = 1:1: 6$;
3. Obtain the number of ON state devices: $N_{i,on} = N_i^h \cdot \tilde{u}_{i,t}$;
4. Form turned-ON list and turned-OFF list according to $N_{i,on}$ and the descending order of indoor temperatures of all HVACs;
5. Update indoor temperatures by Eq. (9);
6. Check if all HVACs are in the temperature boundary;
   1) If one HVAC is below the lower boundary, it is moved from turned-ON list to turned-OFF list, $N_{i,on} = N_{i,on} -1$;
   2) If one HVAC is above the upper boundary, it is moved from turned-OFF list to turned-ON list, $N_{i,on} = N_{i,on} +1$;
7. Repeat Step 5 and Step 6 until satisfying temperature boundary;
8. Calculate the accumulated dispatching error:
   $e_{i,t} = e_{i,t} + (P_{i,t}^H - P^{rated} \cdot N_{i,on})/6$;
9. End $k$;
10. End $t$.

To make the EV charging close to practical situations, we assume individual EVs charge every few days with long charging times until reaching the upper SOC boundary. Based on this assumption, load dispatching for EV is similar to **Algorithm 2**, thus it is not shown here.

### D. Solution Method of the Coupled First Two Levels

The first level and the second level are coupled due to the inter-dependent variables $P_{i,t}^H$, $P_{i,t}^C$ and $\pi_{i,t}^p$. The solution method for such a coupled bilevel model is described below.

*1) MPEC model:* Since the second level is a linear programming problem, the KKT optimality conditions are the necessary and sufficient conditions of its optimal solution. Therefore, the bi-level optimization problem is transformed into a single level problem with the KKT conditions added to the first level. The new model is a single-level MPEC as:

$$\min (23a) \quad (29a)$$

$$s.t. \text{ constraints (23b), (24b) - (24d), (28)} \quad (29b)$$

$$c_{i,t} - \lambda_t^p \left( 1 - \frac{\partial P_t^{loss}}{\partial P_{i,t}^G} \right) + \lambda_t^q \frac{\partial Q_t^{loss}}{\partial P_{i,t}^G} - \sum_{j \in B} \left( \omega_{j,t}^{v,\min} - \omega_{j,t}^{v,\max} \right) Z_{j,i}^p \quad (29c)$$
$$- \omega_{i,t}^{p,\min} + \omega_{i,t}^{p,\max} = 0, \forall i \in G, \forall t \in T$$

$$d_{i,t} - \kappa_{i,t}^- - \kappa_{i,t}^+ = 0, \quad \forall i \in G, \forall t \in T \quad (29d)$$

$$\lambda_t^p \frac{\partial P_t^{loss}}{\partial Q_{i,t}^G} - \lambda_t^q \left( 1 - \frac{\partial Q_t^{loss}}{\partial Q_{i,t}^G} \right) - \sum_{j \in B} \left( \omega_{j,t}^{v,\min} - \omega_{j,t}^{v,\max} \right) Z_{j,i}^q \quad (29e)$$
$$- \omega_{i,t}^{q,\min} + \omega_{i,t}^{q,\max} - \kappa_{i,t}^- + \kappa_{i,t}^+ = 0, \quad \forall i \in G, \forall t \in T$$

$$0 \leq \omega_{j,t}^{v,\min} \perp \left( V_{j,t} - V^{\min} \right) \geq 0, \quad \forall j \in B, \forall t \in T \quad (29f)$$

$$0 \leq \omega_{j,t}^{v,\max} \perp \left( V^{\max} - V_{j,t} \right) \geq 0, \quad \forall j \in B, \forall t \in T \quad (29g)$$

$$0 \leq \omega_{i,t}^{p,\min} \perp \left( P_{i,t}^G - P_{i,t}^{G,\min} \right) \geq 0, \quad \forall i \in G, \forall t \in T \quad (29h)$$

$$0 \leq \omega_{i,t}^{p,\max} \perp \left( P_{i,t}^{G,\max} - P_{i,t}^G \right) \geq 0, \quad \forall i \in G, \forall t \in T \quad (29i)$$

$$0 \leq \omega_{i,t}^{q,\min} \perp \left( Q_{i,t}^G - Q_{i,t}^{G,\min} \right) \geq 0, \quad \forall i \in G, \forall t \in T \quad (29j)$$

$$0 \leq \omega_{i,t}^{q,\max} \perp \left( Q_{i,t}^{G,\max} - Q_{i,t}^G \right) \geq 0, \quad \forall i \in G, \forall t \in T \quad (29k)$$

$$0 \leq \kappa_{i,t}^- \perp \left( \hat{Q}_{i,t}^G + Q_{i,t}^G \right) \geq 0, \quad \forall i \in G, \forall t \in T \quad (29l)$$

$$0 \leq \kappa_{i,t}^+ \perp \left( \hat{Q}_{i,t}^G - Q_{i,t}^G \right) \geq 0, \quad \forall i \in G, \forall t \in T \quad (29m)$$

Noted that (29c), (29j) and (29k) including all DGs are written in the same generic expression for simplicity due to the page limit. They can be further elaborated based on constraints (19) and (21) for different DGs.

$$\sum_{t \in T} \left( \sum_{i \in H} \pi_{i,t}^p \cdot P_{i,t}^H + \sum_{i \in E} \pi_{i,t}^p \cdot P_{i,t}^C \right)$$
$$= \sum_{t \in T} \sum_{i \in H} \left( \lambda_t^p + \lambda_t^p \frac{\partial P_t^{loss}}{\partial P_{i,t}^D} + \lambda_t^q \frac{\partial Q_t^{loss}}{\partial P_{i,t}^D} + \sum_{j \in B} (\omega_{j,t}^{v,\min} - \omega_{j,t}^{v,\max}) Z_{j,i}^p \right) P_{i,t}^H$$
$$+ \sum_{t \in T} \sum_{i \in E} \left( \lambda_t^p + \lambda_t^p \frac{\partial P_t^{loss}}{\partial P_{i,t}^D} + \lambda_t^q \frac{\partial Q_t^{loss}}{\partial P_{i,t}^D} + \sum_{j \in B} (\omega_{j,t}^{v,\min} - \omega_{j,t}^{v,\max}) Z_{j,i}^p \right) P_{i,t}^C$$
$$= \sum_{t \in T} \sum_{i \in G} c_{i,t} \cdot P_{i,t}^G + d_{i,t} \cdot \hat{Q}_{i,t}^G$$
$$- \left\{ \begin{array}{l} \sum_{t \in T} \lambda_t^p \left( \sum_{i \in F} \left( 1 + \frac{\partial P_t^{loss}}{\partial P_{i,t}^D} \right) P_{i,t}^D + \sum_{i \in F} \frac{\partial P_t^{loss}}{\partial Q_{i,t}^D} Q_{i,t}^D + P_t^{loss0} \right) \\ + \sum_{t \in T} \lambda_t^q \left( \sum_{i \in F} \left( 1 + \frac{\partial Q_t^{loss}}{\partial Q_{i,t}^D} \right) Q_{i,t}^D + \sum_{i \in F} \frac{\partial Q_t^{loss}}{\partial P_{i,t}^D} P_{i,t}^D + Q_t^{loss0} \right) \\ + \sum_{t \in T} \sum_{j \in B} \omega_{j,t}^{v,\min} \left( V^{\min} - V_{1,t} + \sum_{i \in F} Z_{j,i}^p P_{i,t}^D + \sum_{i \in F} Z_{j,i}^q Q_{i,t}^D \right) \\ - \sum_{t \in T} \sum_{j \in B} \omega_{j,t}^{v,\max} \left( V^{\max} - V_{1,t} + \sum_{i \in F} Z_{j,i}^p P_{i,t}^D + \sum_{i \in F} Z_{j,i}^q Q_{i,t}^D \right) \\ + \sum_{t \in T} \sum_{i \in G} \left( \omega_{i,t}^{p,\min} P_{i,t}^{G,\min} - \omega_{i,t}^{p,\max} P_{i,t}^{G,\max} \right) \\ + \sum_{t \in T} \sum_{i \in G} \left( \omega_{i,t}^{q,\min} Q_{i,t}^{G,\min} - \omega_{i,t}^{q,\max} Q_{i,t}^{G,\max} \right) \end{array} \right\} \quad (30)$$

*2) MILP model:* From the above formulation, it can be seen that MPEC is a nonlinear model with nonlinearities in two parts: 1) complementary slackness constraints (29f) - (29m); 2) $\pi_{i,t}^p \cdot P_{i,t}^H + \pi_{i,t}^p \cdot P_{i,t}^C$ in the objective function. The methods for dealing with these nonlinearities are described below [25][26].

For $\pi_{i,t}^p \cdot P_{i,t}^H + \pi_{i,t}^p \cdot P_{i,t}^C$, the strong duality theory states that the primal problem and its dual problem have the same optimal value if the problem is convex. The original second level is a linear programming, thus (29a) can be substituted by (30).

For the complementary constraints, the big-M approach [27] is adopted. Then, each formulation of $0 \leq \omega_{i,t} \perp h_i(x) \geq 0$ can be substituted as:

$$0 \leq \omega_{i,t} \leq M_{i,t} v_{i,t}, 0 \leq h_i(x) \leq M_{i,t}(1 - v_{i,t}) \quad (31)$$

where $M_{i,t}$ is a big number and $v_{i,t}$ is a binary variable.

Now, the completed MILP model is presented as:

$$\min (30) \quad (32a)$$
$$s.t. \text{ constraints } (29b) - (29g), (31) \quad (32b)$$

After the MILP model is solved, **Algorithm 2** is applied to dispatch the aggregated load to all end residents, as discussed in the previous subsection III-C.

It should be noted that the proposed tri-level approach is based on a competitive DLMP model, which is aligned with increasing interests in the industrial practices towards competitive distribution or retail markets [28].

## IV. CASE STUDIES

In this section, the proposed tri-level model is tested on a modified IEEE 33-Bus distribution system. Simulations are performed on a laptop with Intel (R) Core™ i7-8650U 2.11GHz CPU, and 16GB RAM. The coding work is carried out in MATLAB R2019a, YALMIP and CPLEX 12.9.

### A. IEEE 33-Bus Distribution System

The topology of the modified IEEE 33-Bus system is shown in **Error! Reference source not found.**, in which two 500 kW PVs are installed at nodes 12 and 28, respectively; two 500 kW MTs are located at nodes 18 and 33, respectively; and three 500 kVar SVCs are installed at nodes 10, 16 and 30, respectively. The parameters of the HVAC aggregators and the EV aggregators are shown in TABLE II.

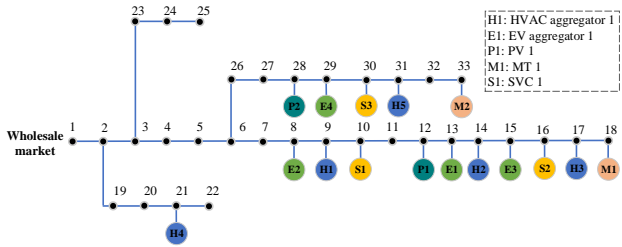

Fig. 2. Modified IEEE 33-Bus system

The day-ahead active LMP of wholesale market, power output of PV, and scaled fixed load are obtained from PJM [29], as depicted in Fig. 3 (a). The system peak fixed load is 3.715MW + j1.78MVar. The active bidding prices of PV and MT are set to $15/MWh and $70/MWh, respectively. Reactive LMP, reactive power price of PV, MT and SVC are set to 0. The outdoor temperature in a summer day is shown in Fig. 3 (b).

Aggregators can choose to participate in the market or not. If they participate in the market, their loads become flexible; otherwise they are only price takers and are regarded as fixed loads. Five cases with different flexible load ratios are established in TABLE III. The flexible load ratio is defined as the maximum ratio of flexible load divided by the system load.

TABLE II. PARAMETERS OF AGGREGATORS

| Aggregators | Parameter | Typical Value |
|---|---|---|
| HVAC Aggregators | Location | Node 9, 11, 17, 21, 31 |
| | Number of HVACs | 160, 160, 320, 320, 320 |
| | $R_A$ (°C/kW) | 3.96, 3.6, 3.96, 4.32, 4.68 |
| | $C_A$ (kWh/°C) | 3.75 |
| | Synchronization limit | [0.1, 0.7] |
| | SOC limit | [0.15, 0.8] |
| | Ramp up/down rate | 0.1 |
| EV Aggregators | Location | Node 8, 13, 15, 29 |
| | Number of EVs | 300, 300, 300, 300 |
| | Number of chargers | 25, 25, 25, 25 |
| | SOC limit | [0.2, 0.8] |

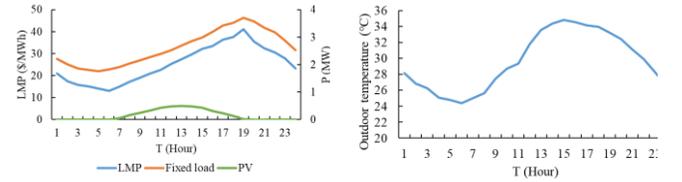

(a) LMP, fixed load and PV profiles  (b) Outdoor temperature
Fig. 3. LMP, fixed load, PV, and outdoor temperature in a summer day

TABLE III. DIFFERENT CASES

| Case | Flexible ratio | Load composition | | |
|---|---|---|---|---|
| | | **Flexible load** | **Total fixed loads** | |
| 0 | 0% | None | H1 - H5, E1 - E4 | Other fixed loads |
| 1 | 10% | H1 - H2, E1 | H3 - H5, E2 - E4 | Other fixed loads |
| 2 | 20% | H1 - H3, E1 - E2 | H4 - H5, E3 - E4 | Other fixed loads |
| 3 | 30% | H1 - H4, E1 - E3 | H5, E4 | Other fixed loads |
| 4 | 40% | H1 - H5, E1 - E4 | None | Other fixed loads |

### B. Results of Case 1

*1) DLMP and voltage profiles:* The DLMP and voltage profiles of all nodes for 24 hours are shown in Fig. 4. It can be seen in Fig. 4 (a) that DLMP increases as the node number increases. This is due to the radial topology and power loss is a component of DLMP. The farther away from Node 1, the larger power loss factor will be. Fig. 4 (b) shows that voltages are maintained within the voltage boundary.

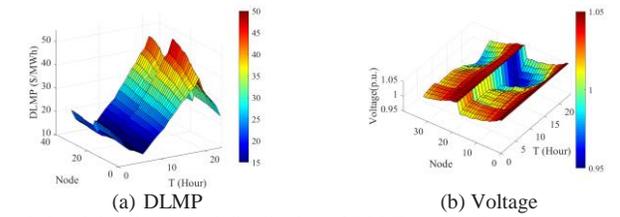

(a) DLMP  (b) Voltage
Fig. 4. Spatial and temporal distribution of DLMPs and voltages

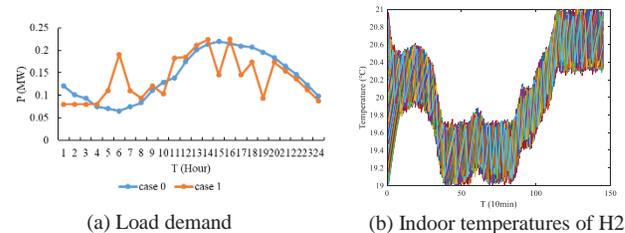

(a) Load demand  (b) Indoor temperatures of H2
Fig. 5. H2's load demand and indoor temperatures

*2) HVAC aggregator:* H2's load demand and its corresponding buildings' indoor temperatures are shown in Fig. 5. From Fig. 5 (a), it can be seen that load shift happens. Combined with Fig. 4 (a), it is revealed that incentivized by DLMP, an aggregator consumes more energy before price arises. According to the indoor temperatures in Fig. 5 (b), this load shifting is the precooling for HVACs.

*3) EV aggregator:* E1's load profile under Case 0 and Case 1 are shown in Fig. 6 (a). In Case 0, we assume that EV owners choose to charge during day time and charging power is evenly distributed at different hours to simplify the charging process. While in Case 1, the EV aggregator makes the optimal charging schedule, such that the charging process happens during 1:00 - 14:00 and 23:00 - 24:00 when DLMP is not high.

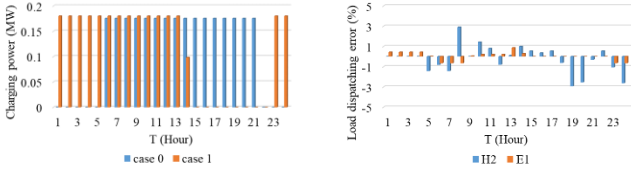

(a) Load profiles of all E1    (b) Load dispatching error
Fig. 6. Load profiles of all E1 and load dispatching error of H2 and E1

The load dispatching errors of H2 and E1 in Case 1 are shown in Fig. 6 (b). The accumulated dispatching error (i.e., $e_{i,t}$ in Step 8, **Algorithm 2**) are within ±3% at each hour, which indicates the effectiveness of the proposed aggregator model and the power dispatch algorithm.

### C. Operation and Economic Analysis under Different Flexible Load Ratios

Case 1 has verified the effectiveness and feasibility of the proposed model and solution method. It also shows load shifting of the flexible load. To have a further understanding of the benefits that the flexible load brings to the entire system, Cases 2-4 have been carried out.

System load profiles under different flexible ratios in 24 hours are shown in Fig. 7, and voltage profiles at $t = 19:00$ are shown in Fig. 8. From Fig. 7 and Fig. 8, it can be observed that with the increase of flexible load ratio, peak load is further reduced, more load is shifted from peak hours to off-peak hours, and the voltage profiles at peak hours are improved.

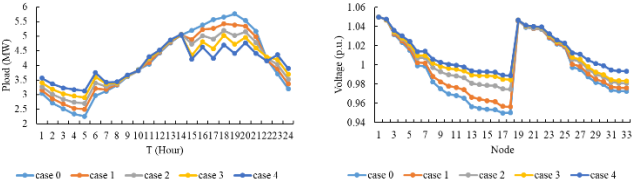

Fig. 7. System load profiles under different flexible load ratios

Fig. 8. Voltage profiles of at $t = 19:00$ under different flexible ratios

TABLE IV. BENEFITS OF DLMP TO THE SYSTEM

| Case | Operation | | Economy | | | | |
|------|-----------|---|---------|---|---|---|---|
|      | $P^*$ (MW) | $V^*$ (p.u.) | GC ($) | HVAC and EV Payment ($) | | | |
|      |           |   |         | Total | Enr. | Vol. | Loss |
| 0 | 5.76 | 0.950 | 2699.9 | 1176.2 | 980.1 | 40.8 | 155.3 |
| 1 | 5.39 | 0.957 | 2629.9 | 1039.9 | 932.4 | 0 | 107.5 |
| 2 | 5.02 | 0.975 | 2580.3 | 971.2 | 900.0 | 0 | 71.2 |
| 3 | 4.72 | 0.985 | 2538.4 | 925.6 | 867.9 | 0 | 57.7 |
| 4 | 4.41 | 0.989 | 2493.3 | 877.4 | 835.2 | 0 | 42.2 |

Part of the numerical results are shown in TABLE IV, in which $P^*$ refers to the system load at $t = 19:00$; $V^*$ refers to the voltage at Node 18 at $t = 19:00$; GC is the generation cost of the DSO; Enr., Vol., and Loss are energy cost, voltage support cost and power loss cost of HVAC and EV payment, respectively. It can be seen that peak load, generation cost, total HVAC and EV electricity payment are reduced by 23.44%, 7.65% and 25.32%, respectively, when Case 4 is compared with Case 0. Due to load shift, the DSO reduces the cost of purchased electricity from the wholesale market and the cost of scheduling peaking DG units (MT in this paper) at high LMP hours. The analysis of cost reduction is presented in detail in subsection IV-D.

### D. Economic Analysis under Different System Load Levels

In this subsection, simulations are carried out to investigate the impacts of the system load level. The fixed load is varied from 0.6 to 1.4 times of the original fixed load shown in Fig. 3 (a). The numbers of HVACs and EVs are kept the same as in TABLE II. The system load level is defined as the peak load in the day of Case 0. As illustrated in Fig. 9, the system load level ranges from 4.3 MW to 7.3 MW, the flexible load of the five cases is between 0 and 2.2 MW.

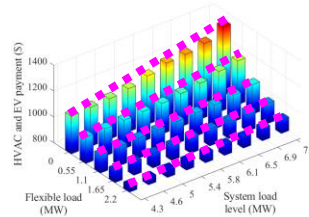 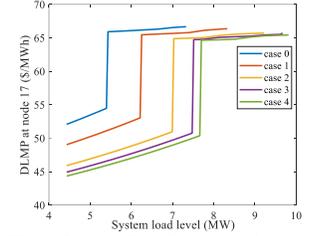

Fig. 9. HVAC and EV payment versus different load levels with various flexible loads

Fig. 10. DLMP with step changes under different system load levels

From Fig. 9, we have three findings regarding the economic benefits from demand flexibility:

- **F1**: At the same system load level, a higher flexible load can significantly reduce the payment which has the same pattern as in TABLE IV.
- **F2**: With the same flexible load, HVACs and EVs payment increases as the system total load increases (i.e., fixed load increases).
- **F3**: In addition to **F2**, the increased payments between 4.3 and 7.3 MW system load levels are different for five cases. The increasing rates are 21.53%, 13.20%, 9.18%, 6.05% and 5.51%, respectively, which indicate a higher flexible load can slow down the payment increase rate. This is shown by the slopes of the five dashed lines in Fig. 9. The detailed increment can be found in TABLE V.

TABLE V. PAYMENT INCREASE AND ITS DECOMPOSITION BETWEEN 4.3 AND 7.3 MW SYSTEM LOAD LEVEL

| Case | ΔPayment ($) | Inc. ratio (%) | Decomposition ($) | | |
|------|--------------|----------------|-------|-------|-------|
|      |              |                | ΔEnr. | ΔVol. | ΔLoss |
| 0 | 235.64 | 21.53% | 0.00 | 147.86 | 87.76 |
| 1 | 132.77 | 13.20% | -0.74 | 62.19 | 71.32 |
| 2 | 86.64 | 9.18% | 4.03 | 22.99 | 59.63 |
| 3 | 54.47 | 6.05% | 0.72 | 0 | 53.75 |
| 4 | 47.13 | 5.51% | -0.58 | 0 | 47.70 |

As Eq. (28) shows, DLMP consists of three parts: energy price, voltage support price and power loss price. Accordingly, HVAC and EV electricity payments can also be decomposed into three components as shown in TABLE IV and TABLE V. Based on observations from these two tables, fundamental reasons of the three findings can be summarized as follows.

- **Load shifts**: In general, if we shift loads from peak DLMP hours to off-peak DLMP hours, the energy cost can be reduced. TABLE IV and the discussion in subsection IV-C clearly demonstrate that with more flexible loads, the load-shifting quantity increases and energy cost decreases. This is an important reason of the finding **F1**. Also, with the same flexible load and an increasing total system load, the load-shifting quantity keeps almost the same. As can be found in TABLE V, ΔEnr. is close to 0 for five cases, thus energy cost contributes little to the for **F2** and **F3**.

- **DLMP step changes**: To illustrate this, Fig. 10 shows the DLMP on Node 17 at $t = 19:00$ under different system load levels. The simulation for each case ends at its own operation capacity. It can be clearly seen that DLMP spikes up at a certain load level. In this study, the step change of DLMP is similar to that of LMP in [30]. However, the difference is that the former is caused by the binding voltage limit while the latter is the binding congestion limit. Taking Case 0 as an example, when the system load is beyond 5.3 MW, DLMP has a sudden increase because the voltage limit at Node 18 becomes binding and M1 is activated to support the voltage near Node 18. The comparison in TABLE IV and TABLE V clearly shows that at the same system load level, the more flexible load, the less likely we have binding voltage limits causing DLMP step changes, thus less voltage support cost. More important, when the system load increases from 4.3 to 7.3 MW, Cases 0, 1 & 2 have a sharper increase than Cases 3 & 4 because DLMP step changes of Cases 0-2 come quicker as shown in Fig. 10. This DLMP step change is a hidden reason related to all the findings **F1~F3**.

- **Power losses**: As shown in Fig. 10, before reaching the step change, DLMP has a slight but consistent increase which is because of power losses. Eq. (28) shows that the power loss price is proportional to power loss factors, which are generally higher in peak hours than that in off-peak hours. Therefore, the loss prices in peak hours are higher than in off-peak hours, and the power loss cost can be reduced by load shift as well. All three findings **F1~F3** are related to power losses.

The above three reasons can serve as general guidelines for decision makers to carefully choose appropriate time and location of HVACs and EVs to potentially give high benefits.

## V. CONCLUSIONS

In this paper, DLMP is used as price signals to guide the electricity consumption of residential customers. The novelty of this paper can be summarized as follows:

1) The proposed HVAC and EV aggregator models can well represent individuals, and the data-driven least squares estimation can model HVACs of different parameters.
2) A tri-level scheduling model is proposed for an aggregator to schedule HVACs and EVs under competitive DLMP paradigm.
3) Three generalized reasons leading to the benefit from flexible HVAC and EV demands are summarized, namely, load shifts, DLMP step changes, and power losses for better decision making.